\documentclass[twocolumn,prb,superscriptaddress,showpacs,preprintnumbers]{revtex4}

\usepackage{amsmath,dcolumn,graphicx}

\begin{document}

\title{
Ytterbium divalency and lattice disorder in near-zero thermal expansion YbGaGe}

\author{C. H. Booth} 
\email{chbooth@lbl.gov}
\affiliation{Chemical Sciences Division, Lawrence
Berkeley National Laboratory, Berkeley, California 94720, USA}

\author{A. D. Christianson}
\altaffiliation[Current address: ]{Center for Neutron Scattering, 
Oak Ridge National Laboratory, Oak Ridge, Tennessee 37831, USA}
\affiliation{Department of Physics and Astronomy, University of California, 
Irvine, California 92697-4575, USA}

\author{J. M. Lawrence}
\affiliation{Department of Physics and Astronomy, University of California, 
Irvine, California 92697-4575, USA}

\author{L. D. Pham}
\affiliation{Physics Department, University of California, 
Davis, California 95616, USA}

\author{J. C. Lashley}
\affiliation{Materials Science Division, Los Alamos National Laboratory, Los Alamos, New Mexico 87545, USA}

\author{F. R. Drymiotis}
\affiliation{Department of Physics and Astronomy,
Clemson University, 
Clemson, South Carolina 29634, USA}

\date{as resubmitted 10/10/6, v2.1}

\preprint{LBNL-60117}

\begin{abstract}
While near-zero thermal expansion (NZTE) in YbGaGe is 
sensitive to stoichiometry and
defect concentration, the NZTE mechanism remains elusive.
We present x-ray absorption spectra that show unequivocally
that Yb is nearly divalent in YbGaGe and the valence does not change with
temperature or with nominally 1\% B or 5\% C impurities, ruling out a valence-fluctuation mechanism.
Moreover, substantial changes occur in the local structure around Yb 
with B and C inclusion.  Together with inelastic neutron
scattering measurements, these data indicate a strong tendency for
the lattice to disorder, providing a possible explanation for NZTE in YbGaGe. 
\end{abstract}



\maketitle

\begin{figure}[b]
\includegraphics[height=3.0in,angle=-90,trim= 170 095 195 95,clip=]{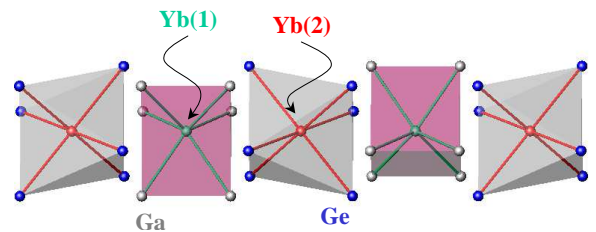}
\caption{
Stacking of the Yb(1)-Ga$_6$ trigonal prisms and the Yb(2)-Ge$_6$ octahedra
in the $P6_3/mmc$ hexagonal cell.
}
\label{struc}
\end{figure}

Observations of near-zero thermal expansion (NZTE) 
are relatively rare, especially near room temperature. The most famous 
example is invar (Fe$_{64}$Ni$_{36}$) which has a volume 
expansion coefficient $\beta \approx 4\times10^{-6}$~K$^{-1}$. Recently, YbGaGe 
(Fig. \ref{struc}) has 
been identified as a potential NZTE material,\cite{Salvador03}
with $\beta \approx -3 \times 10^{-7}$~K$^{-1}$,\cite{note_ybgage} 
depending on the Ga/Ge ratio. In fact, the exact stoichiometry 
is an important factor, with many subsequent measurements unable to reproduce
the original work.\cite{Margadonna04,Muro04,Bobev04,Tsujii05,Drymiotis05} 
In particular,
$\beta$ varies widely between about $4.5 \times10^{-5}$~K$^{-1}$ and 
$1.4 \times10^{-4}$~K$^{-1}$, typical values for intermetallics.
Understanding the differences between 
these samples should eventually allow the reproducible fabrication of, what
should be, a technologically important material.

A fundamental difference between the original and subsequent measurements is
the magnitude and the sign of the magnetic susceptibility. In 
Ref.~\onlinecite{Salvador03} a Curie-Weiss-like susceptibility was observed
above about 100~K with an effective moment of 
$\mu_\textrm{eff} \approx 4.12 \mu_\textrm{B}$, close to the
full moment expected from free Yb$^{3+}$ ions (4.54 $\mu_\textrm{B}$). Below
100~K, a sharp decline in the moment occurred, with 
$\mu_\textrm{eff} \approx 0.82 \mu_\textrm{B}$. These authors pointed out that a
bond-valence sum\cite{Brown85} for each of the two ytterbium
sites (Fig. \ref{struc}) gives a valence of +2.6 for the Yb(1) site and
+2.0 for the Yb(2) site. Such a result is consistent with a
mixed valent state for at least one of the Yb sites. Together with the
magnetic susceptibility data, this observation led the authors to conclude that
the mechanism for the observed NZTE was a change in the Yb valence with 
temperature toward a divalent state below 100~K, consistent with the susceptibility.

In direct contrast to those first measurements, subsequent measurements find a 
diamagnetic susceptibility for the pure 
compound,\cite{Muro04,Bobev04,Tsujii05,Drymiotis05} even over a wide range of
the Ga/Ge occupancy ratio\cite{Tsujii05} and with other 
defects.\cite{Drymiotis05}
These susceptibility measurements are a very strong indication of divalent
Yb at all measured temperatures.
However, if the real Yb valence is +2.6 or even less, this could be an
indication of a very high Kondo temperature, $T_\textrm{K}$, perhaps well in 
excess of 1000~K.  Given this possibility and 
the fact that the magnetic susceptibility is expected to go to a constant 
$\chi_0 \propto 1/T_\textrm{K}$ at temperatures below $T_\textrm{K}$, the
diamagnetic contribution from the nonmagnetic matrix needs to be carefully
considered before declaring that all the Yb is divalent.  
A spectroscopic measurement of the Yb valence would provide a direct
measurement.

As a possible explanation for the apparently anomalous Salvador \textit{et al.} 
work, Drymiotis \textit{et al.}\cite{Drymiotis05} considered
the possibility that the observed NZTE could be a consequence of
disorder. They note that the original synthesis was performed in graphite 
crucibles and therefore grew YbGaGe with small amounts of carbon and boron in 
alumina crucibles.  They found that the thermal expansion could be reduced
by about 50\% with only 0.5~at.\% carbon or boron included in the starting 
materials. (All subsequent concentration percentages are in at.\%). They 
conjecture that, although they did not obtain NZTE and still
observe diamagnetic behavior, that the mechanism for NZTE in 
Ref.~\onlinecite{Salvador03} has a similar origin.

\begin{figure}[t]
\includegraphics[width=3.25in,trim=0 10 0 10,clip=]{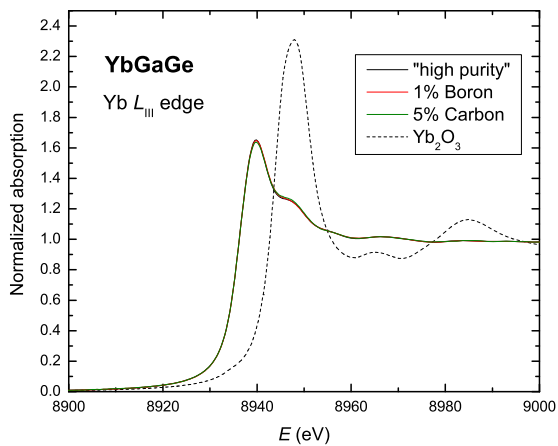}
\caption{
Yb $L_\textrm{III}$-edge XANES spectra for the pure, 1\% B, and the 5\% C
samples of YbGaGe, all collected at 30 K, together with a room-temperature
spectrum of Yb$_2$O$_3$. The three YbGaGe spectra are nearly identical, and
are therefore difficult to discern in this figure.
}
\label{xanes}
\end{figure}

Here, we report x-ray absorption near-edge structure (XANES) measurements 
at the Yb $L_\textrm{III}$ edge that show ytterbium in these materials is, 
in fact,
nearly divalent, that the valence does not change dramatically with carbon
or boron doping, and that the remaining trivalent component is easily explained 
as due to an impurity phase (for example, Yb$_2$O$_3$ or a Yb-Ga binary alloy).
In addition, extended x-ray absorption fine-structure (EXAFS) measurements of 
the local structure environment around the Yb atoms shows that small amounts of 
carbon or boron in YbGaGe have a surprisingly large effect on the average local 
structure. This extreme sensitivity to interstitials appears to be the root 
cause of near-zero thermal expansion in this system.



Samples were prepared as in Ref.~\onlinecite{Drymiotis05}. For the present
work, in addition to a high purity sample grown in a yttria crucible, 
nominally 1\% boron and 5\% carbon doped samples were also fabricated.
X-ray diffraction measurements indicate the YPtAs
structure type in the hexagonal $P6_3/mmc$ space group, with small ($<$5\%)
unidentified impurity phases.  The 
yttria grown sample has $\beta\approx4.6(1) \times10^{-5}$~K$^{-1}$ at room 
temperature according to powder diffraction measurements.\cite{Drymiotis05} 
We have re-arcmelted the doped samples to obtain a polycrystalline average and 
obtained $\beta=3.3(3) \times10^{-5}$~K$^{-1}$ for the 1\% B sample and 
$4.4(2) \times10^{-5}$~K$^{-1}$ for the 5\% carbon sample using a capacitance
dilatometer.\cite{EPAPS_ybgage} 
These values are comparable to the values obtained with diffraction for the
0.5\% B and C doped samples
($3.0(3)\times10^{-5}$~K$^{-1}$ and $3.2(1)\times10^{-5}$~K$^{-1}$, 
respectively\cite{Drymiotis05}). Although the doped samples have been 
re-arcmelted and are compared using different measuring techniques, these data 
counter the notion that more boron or carbon would decrease $\beta$.
Magnetic susceptibility measurements indicate diamagnetic behavior for
all samples with a small impurity tail corresponding to free Yb$^{3+}$ moments.
The effective moment per mole and an estimate of the crystal field effects
indicates the Yb$^{3+}$ impurity fraction is between 1-5\% and shows no 
correlation with B or C concentration.\cite{Drymiotis05,EPAPS_ybgage}

XANES data were collected on Beamline 10-2 at the Stanford Synchrotron 
Radiation 
Laboratory (SSRL).  The samples were first re-ground, passed through a 20 $\mu$m sieve,
brushed onto adhesive tape, and stacked to achieve an absorption length change 
at the Yb $L_\textrm{III}$ edge (8.944 eV) of $\Delta \mu t \approx 1$.
The samples were then placed in a LHe flow cryostat and measured between 30 and 300 K.
Data were collected in transmission mode with a defocused beam from a
half-tuned Si(220) double-crystal monochromator resulting in an energy 
resolution of about 2.0 eV.  Since the core-hole lifetime is about 4.2 eV, the
reported spectra are not lifetime limited. All spectra were 
collected under the same conditions.  
Data were reduced and fit using standard procedures.\cite{Li95b} In particular,
the absorption from the Yb $L_\textrm{III}$ edge as a function of the incident 
energy $E$, $\mu_\textrm{a}(E)$, was isolated from the total absorption 
$\mu(E)$ after approximating all other contributions by extrapolating from the 
pre-edge data and forcing $\mu_\textrm{a}(E)$ above the edge to follow a 
Victoreen formula.\cite{Li95b} The EXAFS oscillations were isolated by
fitting a 7-knot cubic spline to approximate the embedded-atom absorption
$\mu_0(E)$, resulting in the EXAFS function 
$\chi(k)=(\mu_\textrm{a}(k)-\mu_0(k))/\mu_{0}(k)$, where $k$ is the 
photoelectron wave vector obtained from $E$ and the threshold energy $E_0$, 
arbitrarily taken as the energy at the half-height of the edge. The 
oscillations in $\chi(k)$ are due to interference of the back scattered and 
outgoing parts of the photoelectron wave function modulating the absorption
coefficient. A Fourier transform (FT) of $\chi(k)$ produces peaks 
in $r$-space corresponding to scattering shells around the absorbing atomic 
species. Note that phase shifts of the outgoing and back scattered photoelectron
make the FT more 
complicated than a true radial distribution function, so detailed fits using the RSXAP 
package\cite{Hayes82,Li95b,RSXAP} and theoretical lineshapes calculated by 
FEFF7\cite{FEFF6} were performed to extract the local structure information.


The XANES results are displayed in Fig. \ref{xanes}, together with the
spectra of Yb$_2$O$_3$ as a trivalent Yb reference. The main peak
(``white line'') position
of the Yb$_2$O$_3$ reference is typical of trivalent Yb, including in 
intermetallic compounds.  There is a clear energy shift of about 8~eV between 
the YbGaGe compounds and Yb$_2$O$_3$, indicating Yb in these intermetallics 
is predominantly divalent. In addition, we observe a small bump at the 
Yb$_2$O$_3$ white line position, indicating a small trivalent 
component. By fitting these features to a pseudo-Voigt 
function each for the divalent and the trivalent resonances and an integrated 
pseudo-Voigt to model the edge step, we obtain the $f$-hole occupancy 
$n_f=0.17\pm0.10$. 
The small trivalent component could be due to an intermediate valent ground 
state in YbGaGe, but is more likely due to intermetallic or oxide impurities. 
Very little difference in the $f$-occupancy is observed between the three 
samples. Moreover, there is very little change in $n_f$ with temperature, less 
than 1\% between 30 and 300 K.
 

\begin{figure}[t]
\includegraphics[width=3.25in,trim=0 10 0 10,clip=]{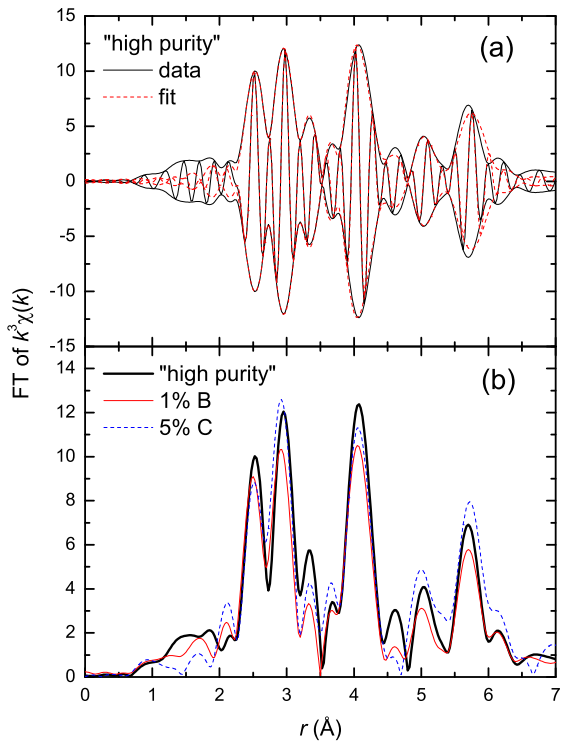}
\caption{
EXAFS data in $r$-space for (a) the pure YbGaGe sample, and (b) all
three samples at 30 K.  The Fourier transforms (FT) are between
2.5 and 16.0 \AA$^{-1}$, Gaussian narrowed by 0.3 \AA$^{-1}$.  The fit in
(a) is between 1.2 and 6.0 \AA and the outer envelope is $\pm$ the 
amplitude, and the modulating inner line gives the real part of the complex 
transform. Only the amplitude is shown in (b).
}
\label{exafs}
\end{figure}

The EXAFS data and fit for the pure YbGaGe sample are shown in 
Fig.~\ref{exafs}.\cite{EPAPS_ybgage}
The data quality is very high and the $r$-space fit quality is also very
good above about 2 \AA. However, below 2 \AA, there
is some unexplained amplitude, as reflected in the reported
$R(\%)=13.1$ (a high-quality EXAFS fit should be near 5\%). Some of this 
amplitude can be explained with between 5-10\% Yb$_2$O$_3$. 
Even so, the fit shown in Fig. \ref{exafs}(a) 
already includes such an impurity, so there is likely some other kind of 
impurity present, possibly in addition to the sesquioxide.  We have tried 
including other intermetallic phases in the fit, such as YbGa and YbGa$_2$, 
with only moderate success.
In any case, the fit parameters are very close to
the expected values from bulk diffraction studies, and the variance of
the pair-distance distribution widths (Debye-Waller factors), $\sigma^2$, are 
also all of a reasonable magnitude.

The $r$-space transforms for all samples at 30 K are shown in 
Fig. \ref{exafs}(b). There are clear differences between the pure and the
substituted samples at all length scales.  The fit results\cite{EPAPS_ybgage} 
for the substituted 
samples are similar to the pure case with the following exceptions:
(1) the bond lengths are very similar except for the Yb(1)-Ge pair in the
1\% boron substituted sample, nominally
at 3.35 \AA, but instead at 3.13(3) \AA; (2) many of the Debye-Waller factors are 
substantially different; and (3) the amplitude reduction factor $S_0^2$ is
unphysically small, $\approx$ 0.5-0.6 in both cases. These results point
strongly to the presence of multiple phases over and above that observed for the
pure YbGaGe sample.  For instance, the short value for the Yb(1)-Ge pair may
be indicative of a YbGa-like phase.\cite{Palenzona79} More directly, the 
reduction in $S_0^2$
(an overall scale factor) and the variations in the Debye-Waller factors,
while still allowing for a high quality fit, can be explained by allowing for
nearly amorphous or highly distorted regions of the sample. The presence of
such a high fraction of sample being so disordered (about 20\%) indicates
that the presence of boron or carbon in the lattice has relatively long-range
effects on the local structure. This result could be verified through 
quantitative analysis of x-ray or neutron diffraction data.

\begin{figure}
\includegraphics[width=3.25in,trim=0 10 0 10]{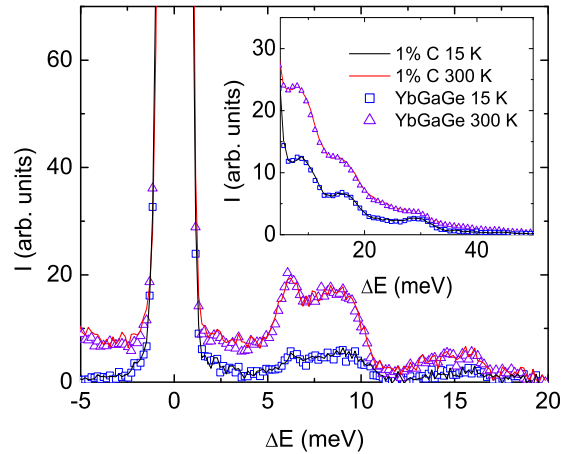}
\caption{Neutron scattering intensity versus energy transfer $\Delta E$ at
15~K and 300~K for undoped and C-doped YbGaGe. The 78~K data are not shown for
clarity. The incident energy in the main 
panel was 25~meV, and 125~meV in the inset.
}
\label{ins}
\end{figure}

In order to determine whether these changes are vibrational in origin, 
inelastic neutron scattering measurements were performed on polycrystalline samples 
of YbGaGe, both undoped and doped with 1\% carbon.  Data were collected 
on the PHAROS time-of-flight spectrometer at the Manuel Lujan Los Alamos 
Neutron Science Center (LANSCE) with incident energies of 25 and 125~meV and 
at $T=$15, 78 and 300 K.  Groups of detectors were summed to improve 
statistics. Figure \ref{ins} shows data from detectors with an average 
scattering angle of θ = 89$^\circ$; for such large scattering angles, the 
spectrum of a polycrystalline sample approximates the phonon density-of-states. 
Peaks are observed near 6, 8.5, 16 and 30~meV.  The fact that the phonon spectra
for both pure and carbon-doped YbGaGe samples are identical means that the 
changes in the thermal expansion and EXAFS data from the C-doped samples 
cannot arise from a bulk distortion of the lattice, since this should affect the
phonon frequencies.  Instead, these changes must be due to random or 
near-random disorder.  Presumably, this disorder reduces the anharmonicity in 
the pair potentials that generate the thermal expansion.  Apparently, these
neutron measurements are not sensitive to such an effect.  Moreover, fits 
to the EXAFS including an anharmonic term, that is, a third 
cumulant,\cite{Eisenberger79} are inconclusive.

These results strongly support the conclusion by 
Drymiotis \textit{et al.}\cite{Drymiotis05} that the mechanism for the reduced
thermal expansion in B and C doped samples, and possibly for the NZTE observed 
by Salvador \textit{et al.},\cite{Salvador03} is chemical defects.  It also 
spectroscopically confirms the conclusions drawn by several 
researchers\cite{Muro04,Bobev04,Tsujii05,Drymiotis05} that the diamagnetic 
susceptibility is indicative of a divalent Yb state at all measured 
temperatures.

Unexplained facts remain regarding the original 
Salvador \textit{et al.}\cite{Salvador03} work. First,
no subsequent published measurement has produced a 
magnetic susceptibility consistent with mostly Yb$^{3+}$
above 100~K. This discrepancy may be due to the exact impurities that
may have been present in the original samples. However, given the 
magnitude of the susceptibility, it seems unlikely that an impurity could 
generate a susceptibility corresponding to bulk Yb$^{3+}$ at room temperature. 
Given this structure's propensity to disorder with small defect concentrations,
the possibility remains that the original samples did, in fact, undergo a
valence change upon cooling.  

Another peculiar fact is that
the bond-valence sums give a much higher valence to the Yb(1) site (+2.6) than
the Yb(2) site (+2.0).  This analysis depends on the $z$-parameter for the Ga
and Ge sites, a quantity that has not been re-measured. The EXAFS results
reported here are similarly insensitive to the differences between the
Ga and Ge atoms as are the x-ray diffraction measurements; however, the results 
are consistent with
two separate Yb sites with different coordination environments, as expected
from the diffraction measurements. Indeed, the local measured bond lengths are 
all consistent with those expected from the diffraction results and their 
associated $z$-parameters.  The bond-valence sum conclusion also 
depends on the relative mixture of Ge and Ga on their nominal sites, a quantity
that is essentially assumed both by the diffraction and the EXAFS measurement.


In conclusion, the $f$-hole occupancy $n_f$ has been measured in YbGaGe using Yb 
$L_\textrm{III}$-edge XANES spectroscopy, and apart from a
small trivalent component that is likely due to oxide or intermetallic 
impurities, Yb in this system appears to be divalent, consistent with
several previous measurements, such as magnetic 
susceptibility.\cite{Muro04,Bobev04,Tsujii05,Drymiotis05} The valence does not 
change dramatically either with temperature or with small amounts of boron or 
carbon substitution. Although the local structure obtained by fitting 
the EXAFS is consistent with that expected from the bulk crystal 
structure, systematic differences with the fitting model indicate the presence
of some other phase, especially in the substituted materials. In particular,
the 1\% boron and 5\% carbon substituted samples
show large differences in the EXAFS compared to the pure sample, indicating
a very large effect on the local lattice for such small amounts of impurities.
The inferred presence of such highly disordered regions is
consistent with the conclusion reached in Ref. ~\onlinecite{Drymiotis05} that 
the reduction of the thermal expansion is related to defects caused by chemical 
substitution.


This work was supported by the Director, Office of Science, Office of Basic 
Energy Sciences (OBES), Chemical Sciences, Geosciences and Biosciences 
Division, U.S.  Department of Energy (DOE) under Contract No. AC03-76SF00098. 
Work at Irvine was supported by DOE Award No. DE-FG02-03ER46036.
XANES data were collected at 
SSRL, a national user facility operated by Stanford University of 
behalf of the DOE/OBES.

\bibliographystyle{apsrev}
\bibliography{/home/hahn/chbooth/papers/bib/bibli}

\begin{thebibliography}{15}
\bibitem[{not()}]{note_ybgage}
\bibinfo{note}{$\beta$ is defined as $V^{-1} \partial V/\partial T$, however,
  it is common practice to use $(V_T-V_i)/V_T$ or $(V_T-V_i)/V_i$ instead. For
  direct comparisons, we use the derivative version, and where it is not
  reported as such, we have extracted $\beta$ from the structural data in a
  given reference.}
\end{thebibliography}

\end{document}